\documentstyle[12pt]{article}
\newcommand{\nn}{\nonumber\\}\newcommand{\p}[1]{(\ref{#1})}
\begin{document}
\renewcommand{\thefootnote}{\fnsymbol{footnote}}
\thispagestyle{empty}
\begin{flushright}
{Preprint DFPD/93/TH/46\\
 June 1993}
\end{flushright}
\vspace{0.5cm}
\begin{center}
{\large\bf D=$(0\vert 2)$ Dirac--Maxwell--Einstein Theory  \\
as a Way for Describing Supersymmetric Quartions\footnote{Work
supported in part by the American Physical Society}\\}
\vspace{1cm}
\renewcommand{\thefootnote}{\dagger}
Dmitrij P. Sorokin\footnote{Supported in part by the European Community
Research Program ``Gauge Theories, Applied Supersymmetry and Quantum
Gravity'' under contract CEE-SCI-CT92-0789}
\renewcommand{\thefootnote}{\ddagger}
\footnote{Permanent address: Kharkov Institute of Physics and
Technology, Kharkov, 310108, the Ukraine
e-mail address:
kfti\%kfti.kharkov.ua@relay.ussr.eu.net}\\
\vspace{0.5cm}
{\sl Dipartimento di Fisica ``G. Galilei'', Universit\'a di Padova\\

\smallskip
Instituto Nazionale di Fisica Nucleare, Sezione di Padova\\
Via Marzolo 8, 35131 Padova, Italia\\}
\vspace{0.5cm}
and\\
\vspace{0.5cm}
Dmitrij V. Volkov\\
\vspace{0.5cm}
{\it Kharkov Institute of Physics and Technology}\\
{\it Kharkov, 310108,  the Ukraine}\\
\vspace{1.5cm}
{\bf Abstract}
\end{center}
Drawing an analogy with the Dirac theory of fermions interacting with
electromagnetic and gravitational field we write down supersymmetric
equations of motion and construct a superfield action for
particles with spin $1\over 4$ and $3\over 4$ (quartions), where the
role of quartion momentum in effective (2+1)--dimensional space-time is
played by an abelian gauge superfield propagating in a basic
two-dimensional Grassmann-odd space with a cosmological constant
showing itself as the quartion mass.
  So, the $(0\vert 2)$ (0 even and 2 odd)
dimensional model of quartions interacting with the gauge and
gravitational field manifests itself as an effective (2+1)-dimensional
supersymmetric theory of free quartions.

\setcounter{page}1
\renewcommand{\thefootnote}{a}
\newpage
\section{Introduction}
In modern theoretical physics there are two branches of research being
closely related to fundamental problems of space-time, namely the
twistor program \cite{pen} and supersymmetry \cite{gvz}. Both of them
are based on the fundamental role played by the commuting and
anticommuting spinors in establishing the relationship between quantum
mechanical properties of physical objects and the geometrical structure
of space-time.

In the present paper we suggest another argument for the idea that
spinors are in the foundation of ``everything'' by proposing a
field-theoretical model in $(0\vert 2)$-dimensional Grassmann-odd
spinor space for describing particles with spin ${1\over 4}$ and
${3\over 4}$ called quartions \cite{v}. The counterparts of the
quartions with respect to quantum statistics are known under the name
`semions' \cite{z} (because of the middle position they take between
bosons and fermions). Since the semions are
mainly considered in the Chern--Simons approach to the anyons, which
has not been proved to be completely equivalent to the group-theoretical
approach, and since in the group-theoretical approach the Pauli
spin-statistics principle for anyons has not  been completely proved
yet (though there is a strong evidence \cite{fm} that it does take
place) \footnote{On general problems of group-theoretical
approach to
describing particles with fractional spin and statistics see
\cite{fm,jn,p,f} and references therein.}, we distinguish the
quartions and the semions by their names, and just the quartions are
the subject of the present paper.

The quartions and semions attract great
deal of attention because of their possible relation to the problem of
high-${\rm T}_c$ superconductivity and other problems of strongly
correlated quantum electron systems in two-dimensional space
\cite{a,z,fms},
and due to their peculiar group-theoretical \cite{v,sv} and
interaction \cite{ks} properties. Besides, quartions may provide a
consistent way of describing solitons with the spin values $s={1\over
4},~{3\over 4}$ which were discovered in a $D=2+1$ chiral
\hbox{$\sigma$--model} by Balachandran et. al. \cite{ba}, and one may
presume some
correspondence between this chiral model and a model of interacting
quartions in analogy to the well known equivalence of the sine-Gordon
and the Thirring model \cite{c}.

The development of the group-theoretical approach to anyons is hindered
by a lack of reliable symmetry and geometrical ground, which
results in problems with constructing field equations of motion and
actions for anyons.
Drawing an analogy with the Dirac theory of fermions interacting with
electromagnetic and gravitational field allows us to write down
supersymmetric equations of motion and construct a superfield action
for  quartions, where
the role of quartion momentum in effective (2+1)--dimensional
space-time is played by an abelian gauge superfield propagating in a
basic two-dimensional Grassmann-odd space with a cosmological constant
manifesting itself as the quartion mass. The quartion equations of the
model under consideration coincide with that of an alternative D=2+1
supersymmetric field model  proposed in \cite{sv}. So, the $(0\vert
2)$--dimensional model of quartions interacting with the gauge and
gravitational field manifests itself as the effective (2+1)-dimensional
supersymmetric theory.

\section{Group-theoretical background}
We start to draw the analogy between fermions and quartions by comparing
their Lorentz group representations.
To describe the Lorentz transformations of spinors one introduces the
Dirac matrices $(\gamma^{m})_{\alpha}^{{~}\beta}$ transforming as the
\hbox{D-dimensional} vector representation of the Lorentz group and
forming the Clifford algebra characterized by the anticommutator
\begin{equation}\label{ca}
\{\gamma^{m},\gamma^{n}\}\equiv\gamma^{m} \gamma^{n}+\gamma^{n}
\gamma^{m}=2g^{mn},
\end{equation}
where $g^{mn}$ is the Lorentz metric whose signature is chosen to be
(+,-,...,-). Small latin letters stand for vector indices and small
greek letters stand for spinor indices.

The commutators of the Dirac matrices form the generators of the spinor
representation of the Lorentz group
\begin{equation}\label{spr}
{S^{mn}}={i\over 4}[\gamma^{m},\gamma^{n}].
\end{equation}

{}From the other hand, to describe quartions in D=2+1 one has to consider
infinite-dimensional representations of the Lorentz group $SO(1,2)\sim
SL(2,R)$ characterized by the lowest spin weights ${1\over 4}$ and
${3\over 4}$ \cite{lp}, which may be constructed by means of a Majorana
spinor operator $L^{\alpha}$ ($\alpha =1,2$) \cite{v,sv} forming the
Heisenberg algebra with respect to the commutator
\begin{equation}\label{ha}
[L^{\alpha},L^{\beta}]\equiv{L^{\alpha}L^{\beta}-L^{\beta}L^{\alpha}}
=i\varepsilon^{\alpha\beta},
\end{equation}
where $\varepsilon^{\alpha\beta}=\varepsilon_{\alpha\beta}=
-\varepsilon^{\beta\alpha}$ ($\varepsilon^{12}=1$) is the metric in a
spinor space. An infinite dimensional representation of the
Heisenberg--Weyl group generated by the algebra \p{ha} is well-known
and describes the energy eigenstates of a one-dimensional quantum
oscillator
\begin{equation}\label{o}
|n>={1\over\sqrt{n!}}(a^+)^n|0>\hskip24pt (n=0,1,...,\infty),
\end{equation}
where $a^+={1\over\surd 2}(L_1-iL_2)$ and $a={1\over\surd 2}(L_1+iL_2)$
are the raising and lowering operator, respec\-tively, and $a|0>=0$.

The anticommutators of the $L_\alpha$ components
\begin{equation}\label{ser}
S^{\alpha\beta}={1\over 4}\{L^{\alpha},L^{\beta}\}
\end{equation}
form the infinite-dimensional group representation we are looking for.
Indeed, as one can directly verify using \p{ha}, $S^{\alpha\beta}$
satisfy the commutation relations for the Lorentz generators, and the
Casimir operator ${1\over 2}S_{\alpha\beta}S^{\alpha\beta}$ of the
representation has the eigenvalue $s_0(s_0-1)=-{3\over {16}}$. Thus
$s_0$ is equal to either ${1\over 4}$ or ${3\over 4}$, and with respect
to $SL(2,R)$ the Heisenberg-Weyl group representation \p{o} splits into
the Lorentz group representations with the lowest weights $1\over 4$
and $3\over 4$, which correspond, respectively, to even and odd values
of $n$ in Eq.~\p{o}, both representations being transformed into each
other by $L_\alpha$.

An important assumption we make here is that since the difference
between spin $3\over 4$ and $1\over 4$ is $1\over 2$, the relative
statistics of corresponding quartionic states upon proper quantization
should be fermionic, and, hence, $L_\alpha$ are to be odd operators in
accordance with Eq.~\p{ser}.  So, the algebra ~\p{ser}, with $L_\alpha$
satisfying the additional conditions \p{ha}, generates an infinite
dimensional $OSp(1,2)$ group representation. Note that both the
$OSp(1,2)$ representation and the Heisenberg--Weyl group representation
are realized on the same Hilbert space \p{o}.

We see that $\gamma^m$ and $L^\alpha$ are the antipodes in the sense
that where the commutator arises for the Dirac matrices the
anticommutator arises for $L^\alpha$ and vice versa.
Below we use this interchange in the commuting and anticommuting
properties for constructing the quartion model in analogy with the
Dirac--Maxwell--Einstein theory.

\section{Equations of motion}
Fermions are well-known to be described by spinor wave functions
$\psi_\alpha(x)$ of bosonic space-time coordinates $x^m$. So, there is
a correspondence between $x^m$ and $\gamma^m$ (they both form the
vector representation of the Lorentz group), which is required for
writing down the Dirac equation
\begin{equation}\label{de}
(i\gamma^{m}\partial_{m}+m)\psi(x)=0,
\end{equation}
for free fermions with a mass $m$.

Following the correspondence of the group-theoretical structure of
space-time coordinates to that of the basic operators
\p{ca} and \p{ha} we consider the quartion wave functions
 $\Phi(\theta)$ to be determined in a $(0\vert 2)$--dimensional
 Grassmann-odd space parametrized by anticommuting Majorana spinor
 coordinates $\theta_\alpha$ and transformed as the irreducible
 representation of the Heisenberg-Weyl group \p{o} (or more strictly
 speaking $OSp(1,2)$)
 \begin{equation}\label{sf}
\Phi(\theta) =\sum_{n=0}^{+\infty}\Phi^{(n)}(\theta)|n>.
\end{equation}
Thus $\Phi^{(n)}(\theta)$ are polynomials of the form
 \begin{equation}\label{com}
\Phi^{(n)}(\theta)=A^{(n)}+i\theta^\alpha\psi^{(n)}_{\alpha}+
 i\theta^{\alpha}\theta_{\alpha}C^{(n)}.
 \end{equation}
 Note that $\Phi^{(n)}(\theta)$ are superfields with respect to the
$SL(2,{\bf R})$ rotations and the  shifts
$\theta_{\alpha}\rightarrow\theta_{\alpha}+\epsilon_\alpha$ in the
Grassmann space and have relative fermionic statistics for even and odd
$n$ respectively.

    By analogy with the Dirac equation \p{de} the ``free'' equation
 of motion for quartions is written as
\begin{equation}\label{se}
(iL^{\alpha}{\partial\over\partial\theta^{\alpha}}+\kappa)\Phi(\theta)=0,
\end{equation}
where $\kappa$ is a ``mass'' parameter whose physical meaning will be
determined below.

Next step is to introduce the interaction of the objects in question
 with an abelian gauge field.
In the Dirac--Maxwell theory this is achieved by lengthening the
  derivative in \p{de} with a bosonic vector field $A_m(x)$
described  by the Maxwell action
\begin{equation}\label{maf}
S_{A_m}=-{1\over 4}\int d^{D}xF_{mn}F^{mn}
\end{equation}
(where $F_{mn}=\partial_{m}A_n-\partial_{n}A_m$)
and determined
  up to the gauge transformation
\begin{equation}
A_{m}(x)\rightarrow{A_{m}(x)-\partial_{m}\varphi(x)}, \hskip16pt
  \psi(x)\rightarrow\psi(x)\exp{i\varphi(x)}.
  \end{equation}
Then the Dirac equation takes the well-known form
\begin{equation}\label{die}
(i\gamma^{m}{\cal D}_{m}+m)\psi(x)
\equiv(i\gamma^{m}(\partial_{m}+iA_{m}(x))+m)\psi(x)=0,
\end{equation}
whose integrability condition
\begin{equation}\label{fmm}
({\cal D}_{m}{\cal D}^{m}+
{i\over 2}F_{mn}\gamma^{m}\gamma^{n}+m^2)\psi(x)=0
\end{equation}
indicates
that charged fermions have nonzero magnetic moment (second term in
Eq.~\p{fmm}).

To describe gauge interactions of quartions we introduce real Grassmann
spinor superfield
\begin{equation}\label{a}
A_{\alpha}(\theta)=a_{\alpha}+\theta^{\beta}p_{\beta\alpha}+i\theta^{\beta}
\theta_{\beta}c_{\alpha}
\end{equation}
determined up to  gauge transformations
\begin{equation}\label{at}
A_{\alpha}(\theta)\rightarrow{A_{\alpha}(\theta)-
i\partial_{\alpha}\varphi(\theta)},
\end{equation}
with $\Phi(\theta)$ being transformed as
$$\Phi(\theta)\rightarrow{\Phi(\theta)\exp{i\varphi(\theta)}},$$
where
$\varphi(\theta)=\omega+i\theta^\alpha\rho_\alpha+
i\theta^\alpha\theta_\alpha\chi$.
For $A_{\alpha}$ we wright down the action
\begin{equation}\label{ses}
S_{A_\alpha}=-{1\over 4}\int d^{2}\theta F_{\alpha\beta}F^{\alpha\beta},
\end{equation}
where
$F^{\alpha\beta}(\theta)=\partial_{\alpha}A_{\beta}(\theta)
+\partial_{\beta}A_{\alpha}(\theta)=p_{\alpha\beta}+p_{\beta\alpha}
+2i(\theta_{\alpha}c_{\beta}+\theta_{\beta}c_{\alpha})$.
is the gauge field strength.

One can easily see that $a_\alpha$ and $p_{\beta\alpha}\varepsilon^
{\alpha\beta}$ component of $A_\alpha(\theta)$  may be gauged away;
   hence, $p_{\alpha\beta}$ is a real symmetric matrix
   represented in the form
\begin{equation}\label{p}
p_{\beta\alpha}=ip_m\gamma^{m}_{\alpha\beta},
   \end{equation}
where $p_m$ is a constant vector to be interpreted below as the quartion
momentum in the effective (2+1)--dimensional Minkowski space-time. Note
also that the $c_\alpha$ component of $A_\alpha$ has the meaining of a
quartion current (see next section). So the gauge field turns out to
carry such dynamical characteristics of the quartions as the momentum
and the current.

In the presence of the gauge field $A_{\alpha}(\theta)$ the quartion
equation takes the form
\begin{equation}\label{mas}
(iL^{\alpha}{\cal D}_{\alpha}+\kappa)\Phi(\theta)\equiv
{(iL^{\alpha}(\partial_{\alpha}+
A_{\alpha}(\theta))+\kappa)\Phi(\theta)}=0,
\end{equation}
whose integrability condition
\begin{equation}\label{mms}
({i\over 2}{\cal D}_{\alpha}{\cal D}^{\alpha}+
{1\over 2}F_{\alpha\beta}L^{\alpha}L^{\beta}
+\kappa^2)\Phi(x)=0
\end{equation}
indicates that the quartions possess nonzero ``magnetic'' moment, which
    will be interpreted as the quartion helicity times a quartion mass
in the effective (2+1)--dimensional space-time.

Note the presence of the imaginary unit $i$ in the transformation law
\p{at} for $A_\alpha(\theta)$ and the absence of $i$ in the spinor
derivative \p{mas}. This is because
$\partial\over{\partial\theta^\alpha}$, in contrast to
$\partial\over{\partial x^m}$,  is
a Hermitian operator.

Let us consider now the simplest case corresponding to the propagation
of a quartion wave function in an external gauge field
$A_{\alpha}(\theta)$ satisfying the free equations of motion
\begin{equation}\label{free}
{\partial F^{\alpha\beta}\over{\partial\theta^\alpha}}=0.
\end{equation}
It is easy to verify that the single solution of Eq.~\p{free} is
$c_\alpha=0$; hence, the only physically nontrivial component of the
superfield \p{a} is the vector  $p_{\alpha\beta}$, and \p{mas} and
\p{mms} are reduced to
\begin{equation}\label{ld}
(iL^{\alpha}{\rm D}_{\alpha}+\kappa)\Phi(\theta)=0,
\end{equation}
\begin{equation}\label{ic}
(iL^\alpha{\rm D}_\alpha-\kappa)(iL^\alpha{\rm D}_\alpha+\kappa)\Phi=
({i\over 2}{\rm D}_\alpha{\rm D}^\alpha\Phi(\theta)+
p_{\alpha\beta}L^{\alpha}L^{\beta})-\kappa^{2})\Phi(\theta)=0.
\end{equation}
where in ${\rm
D}_{\alpha}\equiv{\partial\over\partial\theta^{\alpha}}+\theta^{\beta}
p_{\beta\alpha}$ one may recognize the supercovariant derivative of
 the $\hbox{N=1}, D=2+1$ SUSY theory \cite{g} in the momentum
representation, while, due to Eq.~\p{ser}, the ``magnetic moment'' term
$p_{\alpha\beta} L^{\alpha}L^{\beta}$ is nothing but the
Pauli-Lyubanski scalar of the $D=2+1$ super Poincar\`e group   (see,
for example, \cite{sv}).  Indeed, since for $A_{\alpha}(\theta)$ with
$c_\alpha=0$ the following anticommutation relations take  place:
\begin{equation}\label{sa}
\{{\rm D}_\alpha,{\rm D}_\beta\}=2p_{\alpha\beta},\hskip16pt
\{Q_\alpha,Q_\beta\}=-2p_{\alpha\beta},\hskip16pt \{Q_\alpha,{\rm
D}_\beta\}=0,
\end{equation}
(where
$Q_\alpha = {\partial\over{\partial\theta^\alpha}}-A_{\alpha}(\theta)$
   is a supersymmetry generator), it follows that Eq.~\p{ld} is
invariant under supersymmetry transformations
\begin{equation}\label{gst}
\delta\Phi(\theta)=i\epsilon^\alpha Q_\alpha\Phi(\theta).
\end{equation}

Note that when $c_\alpha=0$ the gauge condition resulting in
$p_{\alpha\beta}=p_{\beta\alpha}$ looks as
$\partial_\alpha A^\alpha(\theta)=0$, which is analogous to the Lorentz
condition $\partial_lA^l(x)=0$ for the electromagnetic field.

 As in the case of the Majorana equation \cite{m}, Eq.~\p{ld} describes
an infinite tower of spinning states with tachyon being present in the
 spectrum. To single out quartionic states with positive mass square and
 definite values of helicity one has to impose the constraint
 \begin{equation}\label{msc}
{1\over
 4}F_{\alpha\beta}F^{\alpha\beta}=2m^2=p_{\alpha\beta}p^{\alpha\beta},
 \end{equation}
 which plays the role of the mass shell condition for quartions ($m$ is
 a constant). To find the solutions of Eqs.~\p{ld} and \p{ic}, with
 \p{msc} being taken in to account, let us wright down the
 equations for $\Phi(\theta)$ components which follow from \p{ld} and
 \p{ic}. We get
 \begin{eqnarray}\label{comp}
 L^\alpha\psi_\alpha=\kappa A,\nn
 L^\beta p_{\beta\alpha}A-2iL_\alpha C=\kappa\psi_\alpha,\nn
 {i\over 2}L^\beta p_{\beta\alpha}\psi^\alpha=\kappa C
  \end{eqnarray}
 and
\begin{eqnarray}\label{coi}
2C=(p_{\alpha\beta}L^{\alpha}L^{\beta}-\kappa^{2})A,\nn
{1\over 4}p_{\alpha\beta}p^{\alpha\beta}A=(p_{\alpha\beta}
L^{\alpha}L^{\beta}-\kappa^{2})C, \nn
p_{\alpha\beta}\psi^\beta=
i(p_{\beta\gamma}L^{\gamma}L^{\beta}-\kappa^{2})\psi_\alpha.
\end{eqnarray}

For simplicity we shall perform further consideration of Eqs.~\p{comp}
and \p{coi} in the rest frame
$$p_0^n=(m,0,0),\hskip18pt
(p_0)_\alpha^{~\beta}=im\gamma^{0\beta}_\alpha$$.
{}From Eqs.~\p{coi} and \p{msc}, with taking into account
\begin{equation}\label{plc}
p_{\alpha\beta}L^{\alpha}L^{\beta}\sum_{n=0}^{+\infty}A^{(n)}|n>=
 m\sum_{n=0}^{+\infty}(2n+1)A^{(n)}|n>,
\end{equation}
it follows that
\begin{equation}\label{mk}
((2n+1)m-\kappa^2)^2=m^2,
\end{equation}
and $\kappa^2$ and $m$ are to be connected with each other by the
 equation
\begin{equation}\label{sqr1}
\kappa^2 =2mn,
\end{equation}
 or
 \begin{equation}\label{sqr2}
\kappa^2 =2m(n^\prime+1),
\end{equation}

 It is essential that for given values of $\kappa^2$ and $m$
  Eq.~\p{ic} (or \p{coi}) singles out quartion states characterized by
  the number $n$ or $n^\prime=n-1$. So the only nonzero components of
  $A$ and $C$ are
  \begin{equation}\label{n}
  mA^{(n)}=2C^{(n)},
  \end{equation}
  for $\kappa^2$ determined by \p{sqr1}, and
  \begin{equation}\label{n-1}
  mA^{(n-1)}=2C^{(n-1)},
  \end{equation}
 for $\kappa^2$ determined by \p{sqr2} with $n^\prime=n-1$. In the first
  case (Eqs.~\p{sqr1}, \p{n}) we have for
  $\sum_{n=0}^{+\infty}\psi_\alpha^{(n)}|n>$ the equation
\begin{equation}\label{fc}
ip_\alpha^\beta\psi_\beta^{(n^\prime)}=(2(n^\prime-n)+1)m\psi_\alpha^{(n^\prime)}=0,
\end{equation}
  from which it follows that for \p{fc} to be consistent with \p{msc}
   $n^\prime$ must be equal to either $n$ or $n-1$, the first value being
  excluded by Eqs.~\p{comp}, which indicates that the
  nonzero values of $n$ of quartionic states
  corresponding to the ``scalar'' and ``spinor''
  components of $\Phi(\theta)$ are differed by 1.
  Hence $\psi_\alpha^{(n-1)}$ satisfy the Dirac equation
  \begin{equation}\label{dn-1}
  ip_\alpha^\beta\psi_\beta^{(n-1)}+m\psi_\alpha^{(n-1)}=0,
  \end{equation}
  which leaves nonzero only
  $\psi_{-}^{(n-1)}={1\over\sqrt2}(\psi_{1}^{(n-1)}+i\psi_{2}^{(n-1)})$
  component of $\psi_{\alpha}^{(n-1)};$ $ \psi_{-}^{(n-1)}$ being
  connected with $A^{(n)}$ due to Eq.~\p{comp}:
  \begin{equation}\label{p-a}
  i(-1)^n\psi_{-}^{(n-1)}=\sqrt{2m}A^{(n)}.
  \end{equation}
  Factor $(-1)^n$ appears in \p{p-a} because of the (anti)commuting
  properties of $L_\alpha$ and the corresponding field components.

  Starting with \p{sqr2} and following the same reasoning as above, we
  find the nonzero components of $\Phi(\theta)$ to be
  $A^{(n-1)}$ and $C^{(n-1)}$ which satisfy \p{n} and,
  $\psi_{+}^{(n)}={1\over\sqrt2}(\psi_{1}^{(n)}-i\psi_{2}^{(n)})$
  component of $\psi_\alpha$ which satisfies the Dirac equation
  \begin{equation}\label{dn}
  ip_\alpha^\beta\psi_{\beta}^{(n)}-m\psi_\alpha^{(n)}=0,
  \end{equation}
  and
  \begin{equation}\label{p+a}
  i(-1)^n\psi_{-}^{(n)}=-\sqrt{2m}A^{(n-1)}.
  \end{equation}

  Thus, the general solution of \p{comp} (in the rest frame) is
  \begin{eqnarray}\label{gs}
  \Phi(\theta)=A^{(n)}(|n>+\sqrt{2m}(-1)^n\theta_{+}|n-1>+im\theta_{+}
  \theta_{-}|n>)\nn
  + A^{(n-1)}(|n-1>+\sqrt{2m}(-1)^n\theta_{-}|n>+im\theta_{+}
  \theta_{-}|n-1>)
  \end{eqnarray}
  (no summation over $n$ is implied!).
  It should be noted once again that for $\Phi(\theta)$ to possess
  definite statistics the components and the basic vectors,
  characterized by indices $n$ differed by 1, must have the odd
  relative statistics.

Consider now the case $\kappa=0$. At this
Eq.~\p{mk} has the only one solution \p{sqr1} with $n=0$. The solution
\p{sqr2} should be skipped, since for our choice of the quartion
representations $n$ is the non-negative integer. As a result all
$\Phi(\theta)$ components satisfying \p{comp} correspond to the lowest
state of the representation \p{o}, and $\Phi(\theta)$ obeys the
equations
\begin{equation}\label{som}
L^\alpha{\rm D}_\alpha\Phi(\theta)=0,
\end{equation}
\begin{equation}\label{som1}
p_{\alpha\beta}L^{\alpha}L^{\beta}\Phi(\theta)=m,
\end{equation}
\begin{equation}\label{sca}
({\rm D}_\alpha{\rm D}^\alpha-2im)\Phi(\theta)=0,
\end{equation}
where \p{sca} is nothing but the equation of motion for the massive
scalar superfield in \hbox{$N=1$}, $D=2+1$ superspace.  Eq.~\p{som}
together with \p{sca} reproduces equations of motion for a quartion
superfield with the lowest helicity $1\over 4$ obtained in \cite{sv}.
Thus, the propagation of a two-dimensional quartionic field in the free
gauge field $A_\alpha(\theta)$, that is one described  by Eqs.~\p{free}
and \p{msc}, is equivalent to the propagation of free superparticle
quartionic states with the mass $m$ and the superhelicities ${1\over 4}
+ (n-1)$ and ${1\over 4}+n$ (for $\kappa=2mn\not=0$), and a single free
superparticle quartionic state with the superhelicity $1\over 4$ (for
 $\kappa=0$).

\section{Dirac--Maxwell--Einstein action for quartions}
Let us firstly try to construct a quartion action, from which the
quartion equation of motion \p{mas} can be obtained, as the direct
counterpart of the Dirac--Maxwell theory:
\begin{equation}\label{dta}
S=\int
d^2\theta\bigl(\Phi^\dagger(\theta)(iL^{\alpha}{\cal
D}_{\alpha}+\kappa)\Phi(\theta)-{1\over 4}F_{\alpha\beta}
F^{\alpha\beta}\bigr)
\end{equation}
where $\Phi^\dagger(\theta)$ is Hermitian conjugate to  $\Phi(\theta)$.

Equations of motion for $A_\alpha(\theta)$, which follow from the
action~\p{dta}, have the form
\begin{equation}\label{ae}
{\partial F^{\beta\alpha}\over
 {\partial\theta^\beta}}=i\Phi^\dagger(\theta) L^\alpha\Phi(\theta),
\end{equation}
or
\begin{equation}\label{ce}
c_\alpha={1\over 6}\Phi^\dagger(\theta)L_\alpha\Phi(\theta),
\end{equation}
where $\Phi^\dagger(\theta)L^\alpha\Phi(\theta)$ is a conserved odd
current
analogous to the current $\mathaccent22\psi\gamma_m\psi$ of Dirac
fermions. Just as the Dirac current is nonzero when the fermion is
charged (i.e. described by two irreducible representations of the
Lorentz group) the semion current is nonzero if the both types of the
quartions (with spin $1\over 4$ and $3\over 4$) propagate in the space;
at this $c_\alpha$ component of the gauge field is completely
determined by the quartion current (Eq.~\p{ce}) and does not have
independent degrees of freedom, analogous to the Chern-Simons gauge
field which is completely determined by the current of charged matter
fields \cite{w}. Note that since $c_\alpha$  is a constant independent
of $\theta$ the components of $\Phi^\dagger$ and $\Phi$ should satisfy
the following relations
$${\partial\over{i\partial\theta^\beta}}\Phi^\dagger L_\alpha\Phi=
A^\dagger L_\alpha\psi_\beta-{\psi^\dagger} _\beta L_\alpha A=0,$$
$$({i\over
 2}{\partial\over{\partial\theta_\gamma}}\varepsilon^{\gamma\delta}
{\partial\over{\partial\theta_\delta}})\Phi^\dagger L_\alpha\Phi=
A^\dagger L_\alpha C+C^\dagger L_\alpha A + {i\over 2}{\psi^\dagger} _\beta
L_\alpha\psi^\beta=0,$$
Substituting Eq.~\p{ae} into \p{dta} we get an action for
quartions with ``point-like'' four-quartion interaction
\begin{equation}\label{selfa}
S=\int d^2\theta\Bigl(\Phi^\dagger(\theta)\bigl(iL^{\alpha}
D_{\alpha}+\kappa\bigr)\Phi(\theta)-{1\over
12}\delta(\theta)(\Phi^\dagger L^\alpha\Phi)(\Phi^\dagger
L_\alpha\Phi)\bigr),
\end{equation}
where $\delta(\theta)=\theta_\alpha\theta^\alpha$.

This quartion interaction differs from a current-current interaction of
anyons caused by an induced magnetic moment \cite{l,ks} since the
latter involves anyons with a single value of spin.

The model based on action \p{dta} or \p{selfa} describes interacting
quartions. In general, it contains a tachyon sector since
$p_{\alpha\beta}$ may take space-like values. As we have already seen
above, a possible way to select a physical sector is to restrict the
space of quartion wave functions by the constraint \p{msc} on the gauge
field stress tensor. Eq.~\p{msc} thus imposed is an external constraint
of the model which would be better to obtain from an action as an
equation of motion.

The constraint \p{msc} can be formally introduced
into the action \p{dta} with a corresponding Lagrange multiplier.
It occurs possible to give this multiplier a
geometrical meaning. To this end we generalize \p{dta}
to include gravity in the $(0\vert 2)$--space \cite{av}.

Let us extend the global transformations
$\theta_\alpha\rightarrow\theta_\alpha+\varepsilon_\alpha$ and the
$SL(2,{\bf R})$ rotations of the $\theta_\alpha$ coordinates of the
$(0\vert 2)$--space to general coordinate transformations \cite{av}
\begin{equation}\label{gct}
\theta^{\prime\alpha}=\theta^{\prime\alpha}(\theta).
\end{equation}
These transformations leave invariant the following length element in
the Grassmann-odd space
\begin{equation}\label{le}
g^{\prime}_{\alpha\beta}(\theta^{\prime})d\theta^{\prime\alpha}
d\theta^{\prime\beta}=g_{\alpha\beta}(\theta)d\theta^{\alpha}
d\theta^{\beta},
\end{equation}
where $g_{\alpha\beta}(\theta)$ is an arbitrary antisymmetric metric
which can be represented  in the form
\begin{equation}\label{met}
g_{\alpha\beta}(\theta)={1\over{\rm G}(\theta)}\varepsilon_{\alpha\beta}
\end{equation}
so that ${\rm G}(\theta)$ has the transformation properties of a scalar
density
\begin{equation}\label{tp}
 {\rm G}^\prime (\theta^\prime) =
det({{\partial\theta^{\prime\alpha}}\over {\partial\theta^\beta}}){\rm
 G}(\theta) \end{equation} and plays the role of the gravitation field.
Note that one may fix the gauge, relative to transformations \p{gct},
 in such a way that ${\rm G}(\theta)$ power expansion in
 $\theta^\alpha$ is reduced to \begin{equation}\label{gf} {\rm
G}(\theta)=1+{i\over 2}\theta_\alpha\theta^\alpha{\rm G_0}.
 \end{equation}

The Christoffel symbol and the spin connection satisfying the condition
 of the absence of torsion in the $(0\vert 2)$--space are, respectively
 \begin{equation}\label{29}
 \Gamma^{~~\gamma}_{\alpha\beta}=-\Gamma^{~~\gamma}_{\beta\alpha}=
 \varepsilon^{\gamma\delta}\partial_\delta(\ln
 G)\varepsilon_{\alpha\beta}
\end{equation}
and
\begin{equation}\label{30}
 \omega_{\alpha ab}=\omega_{\alpha ba}= -e^{\beta}_b(\partial_\alpha
 e_{\beta a}-\Gamma^{~~\gamma}_{\alpha\beta}e_{\gamma a})=
-e^{\beta}_b(\partial_\alpha e_{\beta a}+{1\over
G}e^\gamma_a\varepsilon_{\alpha\beta} \partial_\gamma\ln G),
 \end{equation}
 where $e^a_\alpha(\theta)$ are zweinbeins determined by the relations
 $$
 Ge^a_\alpha
 e_{\beta}^b\varepsilon^{\alpha\beta}=\varepsilon^{ab},\qquad
Ge^a_\alpha
 e_{\beta}^b\varepsilon_{ab}=\varepsilon_{\alpha\beta};
 $$
 $a,~b$ are indices of the tangent spinor space.

Taking into account \p{gf} one may choose a local tangent space
 symmetry gauge in such a way that
$e_\alpha^a(\theta)=\delta_{\alpha}^{a} (1-{i\over
 4}\theta_{\beta}\theta^\beta G_0)$.

The curvature tensor is determined as
 $$
 R_{\alpha\beta,\gamma}^{~~~~\delta}=i\bigl(\partial_\alpha
 \Gamma_{\beta\gamma}^{~~\delta}+\partial_\beta
 \Gamma_{\alpha\gamma}^{~~\delta}+\Gamma_{\alpha\gamma}^{~~\rho}
 \Gamma_{\rho\beta}^{~~\delta}+\Gamma_{\beta\gamma}^{~~\rho}
 \Gamma_{\rho\alpha}^{~~\delta}\bigr),
$$
 from which one gets the Ricci tensor and the scalar curvature in the
 form

$$
R_{\alpha\beta}={3i\over 2}{\partial_\gamma\partial^\gamma G
 \over G}\varepsilon_{\alpha\beta}
$$
and
\begin{eqnarray}\label{31}
 R=-GR_{\alpha\beta}\varepsilon^{\beta\alpha}=3i\bigl(\partial_\gamma
 \partial^\gamma G\bigr)=6G_0.
 \end{eqnarray}
Eqs.~(54) indicate that the $(0\vert 2)$--space is an Einstein space of
 a constant curvature \cite{av}.

 In Eq.~\p{31} and below the indices are raised
 and lowered by $\varepsilon^{\alpha\beta}$ and
 $\varepsilon_{\alpha\beta}$, respectively.

 Now, taking into account that the integration measure $d^2\theta$
 transforms under \p{gct} as
 \begin{equation}\label{32}
 d^2\theta^\prime={1\over {\det{({{\partial\theta^{\prime\alpha}}
 \over{\partial\theta^{\beta}}})}}}d^2\theta,
 \end{equation}
 one may write down the generalization of \p{dta} as follows
\begin{equation}\label{33}
 S_G=\int d^2\theta G\Bigl(\Phi^\dagger(\theta)\left(iL^a
e^\alpha_a\nabla_\alpha+\kappa\right)\Phi-R-{1\over
 4}G^2F_{\alpha\beta}F^{\alpha\beta}+6m^2\Bigr),
 \end{equation}
 where
 $\nabla_\alpha={\partial\over{\partial\theta^\alpha}}+{1\over
 4}\omega_{\alpha bc}L^bL^c+ A_\alpha(\theta)$, and $6m^2$ is
 a``cosmological'' constant which should be positive for the model to
 be free of tachyons (see below). Note that this situation is, in some
 sense analogous to that in conventional supergravity theories, where a
 cosmological constant corresponding to the anti--De--Sitter space-time
 is consistent with supersymmetry while a De--Sitter cosmological
 constant is not (see, for example, \cite{n} and references therein).
Let us also remark that in contrast to the ordinary $D=2$ gravity $GR$
 term in \p{33} is {\it not} a total derivative. This is because the odd
 directions imply the ``negative'' dimensionality of the space.

 Varying Eq.~\p{33} over $e^\alpha_a(\theta)$ and converting the result
 with $e^\alpha_a(\theta)$ we get  the Einstein equation in the form
 \begin{equation}\label{34}
2R=6i\partial_\gamma\partial^\gamma G=GT^{~\alpha}_\alpha-3({1\over
 4}G^2F^{\alpha\beta}F_{\alpha\beta}-2m^2),
  \end{equation}
where
\begin{equation}\label{emt}
T_{\alpha\beta}=-T_{\beta\alpha}={i\over
 2}\bigl(\Phi^\dagger(\theta)L^a e_{a[\alpha}
\nabla_{\beta ]}\Phi-\bigl(L^a
 e_{a[\alpha}\nabla_{\beta]}\Phi\bigr)^\dagger\Phi\bigr)=-{\kappa\over
 2G}\varepsilon_{\alpha\beta}\Phi^\dagger\Phi
\end{equation}
 is the quartion energy-momentum tensor.

In the case of $\kappa\not=0$ the solution of Eq.~\p{34} does not
reproduce the solution \p{gs} of \p{ld} and \p{msc} for quartions with
higher helicities, since the matter energy-momentum tensor \p{emt}
contributes to the gravitation equations of motion causing the quartion
dynamics in this case to be more complicated.

For $\kappa=0$ we observe that $T_{\alpha\beta}$ is equal to
 zero on the mass shell (as a consequence of the quartion equations of
 motion). This indicates that quartions with superhelicity ${1\over
 4}$ get nontrivial dynamics only owing to interactions with the gauge
 and gravitational field.  So $T_{\alpha\beta}$ drops from Eq.~\p{34}
 and this allows one to get Eq.~\p{msc} from \p{34}.

In components  (with gauge conditions \p{p} and \p{gf}
imposed, and $\kappa=0$) Eq.~\p{34} looks as
\begin{eqnarray}\label{35}
p_{\alpha\beta}p^{\alpha\beta}=2(m^2-2G_0),\nn
p_{\alpha\beta}c^\beta=0,\nn
3c^\alpha c_\alpha=2i(m^2-2G_0)G_0.
\end{eqnarray}

 The system of Eqs.~(58) has two solutions. The first one is
 $G_0={m^2\over 2}$, $p_{\alpha\beta} p^{\alpha\beta}=0$ (the latter
 being explicitly solved by the Cartan relation
 $p_{\alpha\beta}=\mu_\alpha\mu_\beta$, where $\mu_\alpha$ is a
 commuting Majorana spinor) and $c_\alpha=\varrho\mu_\alpha $ (where
 $\varrho$ is an anticommuting number). This solution is valid also for
 the case of free gravity, when in $(0\vert 2)$--space all matter
 fields are absent, that is it is caused by the gravitational degree
of freedom.  At present time it is not clear whether this solution
leads to interesting consequences for anyon physics.

By dropping the curvature term out of the action we can single out the
second solution of Eqs.~(58) which is just one we are looking for:
 $c_\alpha=G_0=0$, which indicates that on the mass shell the geometry
 of $(0\vert 2)$--space  is flat, $A_\alpha(\theta)$ is a free gauge
 field with $p_{\alpha\beta}$ being timelike, and the cosmological
 constant plays the role of the quartion mass. In the case considered
one may directly verify that Eqs.~\p{som}--\p{sca} for
$\Phi^{({1\over 4})}$ and $\Phi^{({3\over 4})}$ are received from
\p{33} as equations of motion and integrability conditions thereof,
their solutions being $\Phi^{({3\over 4})}=0$ and  the
one-superparticle quartion state with the superhelicity $1/4$ and mass
$m$ for $\Phi^{({1\over4})}$.

\section {Conclusion}
We have constructed the field-theoretical model in the
$(0\vert 2)$--space for describing quartions, which turned out to be
effectively equivalent to the momentum representation of the quartion
field theory in $N=1$, $D=2+1$ superspace. We assume that the action of
the corresponding $D=2+1$ model may be obtained as an effective action
by a functional integration of Eq.~\p{dta} or \p{33} over all possible
independent configurations of $A_\alpha(\theta)$.

We have shown that supersymmetry proved to be intrinsic to the
quartions, since the helicities of the particles with $s={1\over 4}$
and $s={3\over 4}$ are differed by ${1\over 2}(mod~n)$ and their
relative statistics is fermionic.

The model proposed is essentially based on analogy with the
Dirac--Maxwell--Einstein theory, which allowed one to provide all
fields of the model, including the Lagrange multipliers, with clear
geometrical and physical meaning.

Note that this analogy can be drawn even further by starting with the
comparison of the mechanics of a relativistic spinning particle
\cite{b} and that of quartions \cite{sv}. The spin part of the
relativistic spin $1/2$ particle action is described by the term
\begin{equation}\label{spa}
S=\int d\tau i\chi_m{d\over {d\tau}}\chi^m,
\end{equation}
where $\tau$ is a world-line parameter and $\chi^m$ are {\it
anticommuting\/} classical counterparts of the Dirac matrices
$\gamma^m$.  By generalizing Eq.~\p{spa} to be invariant under local
supersymmetry transformations in $(\tau,~\eta)$ world-line
superspace $(\eta^2=0)$ we restore a superfield action for the spin
$1/2$ particle \cite{b}
\begin{equation}\label{sas} S=i\int d\tau d\eta
E(\tau,\eta)\bigl(\partial_\eta+i\eta\partial_\tau\bigr)
X_m\partial_\tau X^m,
 \end{equation}
where the superfield $X^m=x^m(\tau)+i\eta\chi^m(\tau)$ contains
$\chi^m$ and a particle space coordinate $x^m$ as the components.
$E=e(\tau)+i\eta\psi(\tau)$ is the supereinbein which ensures the local
superinvariance of \p{sas}.

By analogy with the spin $1/2$ particle we can write down a spin part
of a classical quartion action in the form \cite{sv}
\begin{equation}\label{ssa}
S=\int d\tau\lambda_\alpha{d\over{d\tau}}\lambda^\alpha,
\end{equation}
where $\lambda^\alpha(\tau)$ are {\it commuting\/} classical
counterparts of $L_\alpha$ \cite{sv}, and get a complete action for a
quartion particle moving in the $(0\vert 2)$--space by generalizing
\p{ssa} to the superfield action
\begin{equation}\label{sss} S=i\int
d\tau d\eta E(\tau,\eta)\bigl(\partial_\eta+i\eta\partial_\tau\bigr)
\Theta_\alpha\partial_\tau\Theta^\alpha,
\end{equation}
where $\Theta_\alpha=\theta_\alpha(\tau)+\eta\lambda_\alpha(\tau) $
contains $\lambda_\alpha$ and $\theta_\alpha$ coordinates of the
quartionic particle.

Adding to \p{sss} a term  $\int d\tau d\eta A_{\alpha}(\Theta)
 (\partial_\eta+i\eta\partial_\tau)\Theta^\alpha$ we get classical
 mechanics counterpart of the model considered.

One may hope that this amusing fermion-quartion analogy, even if being
 a formal trick, may occur to be useful in making deeper
insight into problems of the field-theoretical description of anyons
such as anyon interactions, quantization and spin-statistics
correspondence \cite{fm,jn,p,f}. Moreover, the above simple model
demonstrates that  the role of odd coordinates in describing the real
space-time may be more fundamental then of even ones, the latter being
a manifestation of some fibre bundle structure on a Grassmann base.

\vspace{0.5cm}
{\large\bf Acknowledgments}

\medskip
D.S. is grateful to the European Community for financial support under
the contract CEE-SCI-CT92-0789, the University of Padova and INFN
(Sezione di Padova), and in particular Prof. M. Tonin and Prof. P. A.
Marchetti for kind hospitality
 and valuable discussion.

\medskip
D.V. is grateful to  J.~Lukierski  and S. Majid for stimulative
discussion. The authors are also grateful to I.~Bandos, A.~Pashnev and
V.~Tkach for interest to this work and discussion.

\medskip
The authors gratefully acknowledge the financial support from the
American Physical Society.


\newpage
 \begin{thebibliography}{99}
\bibitem{pen}
{R. Penrose {\sl J. Math. Phys.}
{\bf 8},  345 (1967);\\
R. Penrose and M. A. H. MacCallum, {\sl Phys. Rep.} {\bf
6},  241 (1972)  and refs. therein;\\
R. Penrose, {\sl Rep. Math. Phys.} {\bf
12},  65 (1977).}
\bibitem{gvz}
Y. A. Gol'fand and E. P. Lichtman,
 {\sl Zhurn. Exp. Teor. Fiz.} {\bf 13},  323 (1971);\\
D. V. Volkov  and V. P. Akulov, {\sl JETP Lett.} {\bf 16},  621 (1972);
Sov. Theor.  {\sl Math. Phys.} {\bf 18}  39 (1974);\\
J. Wess  and B. Zumino,  {\sl Nucl. Phys.} {\bf B70},  39 (1974).
\bibitem{v}
D.V.Volkov, {\sl JETP Lett.} {\bf 49}, 473  (1989);\\
D. P. Sorokin, V. I. Tkach and D. V. Volkov in ``Problems of Modern
Quantum Field Theory'' Eds. A. A. Batalin, A. V. Klimyk and
A. B. Zamolodchikov, (Springer-Verlag, 1989).
\bibitem{z}
A. P. Balachandran, E. Ercolessi, G. Morandi, and A. M. Srivastava,
{\sl Int. J. Mod. Phys.} {\bf B4}, 2057 (1990);\\
X. G. Wen and A. Zee, {\sl Phys. Rev.} {\bf B41},  240 (1990);\\
      A.Zee, From semionics to topological fluids. in "Particle
 Physics: VI Jorge Andr\'e Swieca Summer School". Eds. O.J.P.~
      \'Eboli, M.Gomes and A.Santoro (World Scientific, Singapore,
      1992) and refs.  therein.
\bibitem{a}P. W. Anderson, {\sl Science} {\bf 235},  1186 (1988);\\
L. B. Laughlin, {\sl Science} {\bf 242},  525 (1988);\\
L. B. Laughlin, {\sl Phys. Rev. Lett.} {\bf 60}, 2672 (1988).
\bibitem{fms} J. Fr\"ohlih and P. A. Marchetti, {\sl Phys. Rev.}
{\bf B46}, 6535 (1992).
\bibitem{sv} D. P. Sorokin and
D. V. Volkov.  (Anti)commuting spinors and supersymmetric dynamics of
semions.  Preprint IC/92/121, ICTP, Trieste, 1992;
{\sl Nucl. Phys.} {\bf B} (1993) (in press).
\bibitem{ks}J. I. Kogan and G. Semenoff, {\sl Nucl. Phys.} {\bf
B368},  718 (1992).
\bibitem{ba}A. P. Balachandran et.al., {\sl Mod. Phys. Lett.} {\bf
A3}, 1725  (1988).
\bibitem{c} S. Coleman, {\sl Phys. Rev.} {\bf D11}, 2088  (1975).
\bibitem{fm}
J. Fr\"ohlich and P. A.
Marchetti, {\sl Commun. Math. Phys.}  {\bf 121},  177 (1989); {\sl Nucl.
Phys.} {\bf B356}, 533  (1991);\\
P. A. Marchetti, Int. J. Mod. Phys. {\bf B6}, 2275 (1992).
\bibitem{jn}R. Jackiw and V. P. Nair,
{\sl Phys. Rev.} {\bf D43},  1933 (1991).
\bibitem{p}M. S. Plyushchay, {\sl Phys. Lett.} {\bf 248B},  107 (1990);
{\bf 273B}, 250  (1991); {\sl Int. J. Mod. Phys.} {\bf A7},  7045 (1992)
 and refs. therein.
\bibitem{f}
S. Forte, {\sl Rev. Mod. Phys.} {\bf 64}, 193  (1992);\\
R. Iengo and K. Lechner, {\sl Phys. Rep.}, {\bf 213}, 179 (1992).
\bibitem{lp}
S. S. Sannikov, {\sl Ukrainian Phys. Journal} {\bf 10}, 684  (1965);
{\sl Teor.Matem. Fiz.} {\bf 34},  34 (1978);\\
S. Lang, $SL_{2}{\bf R}$ (Springer, Berlin, 1985);\\
A. M. Perelomov, Generalized Coherent States (Nauka, Moscow, 1987).
\bibitem{g}    S. J. Gates, Jr., M. T. Grisaru, M. Roc\^ek and W. Siegel.
       Superspace or one thousand and one lessons in supersymmetry.
       (Benjamin/Cummings Publishing, London, 1983)
\bibitem{m} E. Majorana, {\sl Nuovo Cim.} {\bf 9}, 336  (1932).
\bibitem{w}R. Mackenzie and F. Wilczek, {\sl Int. J. Mod. Phys.} {\bf
A3},  2827 (1988);\\
P. Gerbert, Ibid. {\bf A6}, 173  (1991);\\
F. Wilczek, Fractional statistics and
anyon superconductivity. (World Scientific, Singapore, 1990) and
refs.  therein.
\bibitem{l}S. Latinsky and D. Sorokin, {\sl JETP Lett.} {\bf 53}, 187  (1991);
{\sl Mod. Phys. Lett.} {\bf A38}, 3525  (1991);\\
J. I. Kogan, {\sl Phys. Lett.} {\bf
B262}, 83  (1991).
\bibitem{av} A. P. Akulov and D. V. Volkov, {\sl Teor. Mat. Fiz.} {\bf 41},
147 (1979).
\bibitem{b}L. Brink, S. Deser, B. Zumino, P. di Vecchia and
P. Howe, {\sl Phys. Lett.} {\bf 65B}, 471  (1976).
\bibitem{n} P. van Nieuwenhuizen, {\sl Phys. Rep.} {\bf 68}, 189  (1981).
 \end{thebibliography}
\end{document}